\documentstyle[12pt,epsf]{article}
\begin{document} 
\title{\bf The $B\rightarrow X_sl^+l^-$ and $B\rightarrow X_s \gamma$
   decays with the fourth generation}
 \author{ Chao-Shang Huang,~~~~Wu-Jun Huo, ~~~and Yue-Liang Wu \\ {\sl
Institute of Theoretical Physics, Academia Sinica, P.O. Box $2735$},\\{\sl
 Beijing $100080$, P.R.  China}}

\date{} 
\maketitle

\begin{abstract} 
If the fourth generation fermions exist, the new quarks could influence 
the branching ratios of the decays of
$B\rightarrow X_s \gamma$ and $B\rightarrow X_sl^+l^-$.  We
obtain two solutions of the fourth generation CKM factor $V^{*}_{t^{'}s}V_{t^{'}b}$ from
the decay of $B\rightarrow X_s \gamma$. We use these two solutions to calculate the new contributions of
the fourth generation quark to Wilson coefficients of the  decay of $B\rightarrow X_sl^+l^-$.  
The branching ratio and the forward-backward asymmetry of the decay of $B\rightarrow X_sl^+l^-$ in the
two cases are calculated. Our results are quite different from that of SM in one
case, almost same in another case. If Nature chooses the formmer, the $B$ meson decays could provide a
possible test of the forth generation existence.

\end{abstract}

\section{Introduction}
\label{sec:introduction}

The standard model (SM) is  very successful. But  it is almost certainly incomplete.  For
 example, it does not fix the number of generations.  So far it is not known why
 there is more than one generation and what law of Nature determines their number.  The LEP
 determinations of the invisible partial decay width of the $Z^0$ gauge boson show there are
 certainly three light neutrinos of the usual type with mass than $M_{Z}/2$ \cite{Mark}.
 This result is naturally interpreted to imply that there are exactly three generations of
 quarks and leptons.  However, the existence of the fourth generation is not 
 excluded, if the neutrino of this generation is, for unknown reasons,  heavy, i. e.,
 $m_{\nu_4} \geq M_{Z}/2$ \cite{Berez}. That is, there still exists the room of
 the fourth generation.

If we believe that the fourth generation fermions really exist in Nature, we should give
 their mass spectrums and take into account their physical effects.  In last two decades, many
 theorists have researched this question. 
 For examples, refs.\cite{McKay} researched  the mass spectrum of the fourth generation 
 fermions in the minimal SUSY model and the supergravity model respectlivily.
 Refs. \cite{Swain} considered only the fourth generation neutrino. Refs. \cite{Novi} discussed the limit
 on the masses of the fourth generation neutral and charged leptons,
 $m_{\nu^{'}}$ and $m_{\tau^{'}}$, which had been improved by LEP1.5 to $m_{\nu^{'}}>59$GeV
 and $m_{\tau^{'}}>62$GeV. Ref. \cite{Abachi} reported on 
 a search for pair production of a fourth generation charge $-1/3$ quark $b^{'}$ in 
 $p\bar{p}$ collisions at $\sqrt{s}=1.8$TeV by the ${\rm D\O}$ experiment at the Fermilab 
 Tevatron using an integrated luminosity of 93pb$^{-1}$. There were also many  papers  presented
 other problems about the fourth generation, such as the mass degenercy of the fourth
 generation \cite{Smith}, the vector-like doublet as the fourth generation \cite{Tadashi},
 and so on \cite{others}. Recently, it is noted that the S parmeter measured from precision 
electroweak data is in conflict with a degenerate fourth generation by over three standard 
deviations, or  99.8$\%$ \cite{el}. However, one can get around this discrepancy by assuming 
that there is new physics which partially cancels the contribution of 
fourth generation to the S parameter  (such as additional  Higgs doublets, etc.). 
On the other hand, it is shown that the area of mutual inconsistency between 
the SM and MSSM  Higgs mass bounds is found
to be consistent with the four generation MSSM upper bounds \cite{kang}.  Ref. \cite{Rivero}
 examined the motivation of the existence of generation by operator ordering ambiguity
 and found that there should be four generations. It has been pointed that 
 the fourth generation fermions would increase the Higgs boson 
 production cross section via gluon fusion at hadron collider \cite{Ginz}. The decays of
 the fourth generation fermions  have also been investigated \cite{heh}. The introduction 
of fourth generation fermions can also affect CP violating parameters 
$\epsilon'/\epsilon$ in the kaon system\cite{cwx}.

There were several theoretical schemes to introduce the fourth generation in the SM.
 The most economical and simple one was considered in Ref. \cite{McKay}. The fourth 
 generation model in this note is similar to that.  But we limit ourself to
 the non-SUSY case in order to concentrate on the phenomenological implication of the fourth generation. 
We introduce the fourth generation, an up-like quark $t^{'}$, a
 down-like quark $b^{'}$, a lepton $\tau^{'}$, and a heavy neutrino $\nu^{'}$ in the SM.
 The properties of these new fermions are all the same as their corresponding counterparts 
  of other three generations except their masses and CKM mixing, see tab.1,

\begin{table}[htb] 
\begin{center} 
\begin{tabular}{|c|c|c|c|c|c|c|c|c|} 
\hline 
& up-like quark & down-like quark & charged lepton &neutral lepton \\ 
\hline 
\hline 
& $u$ & $d$& $e$ & $\nu_{e}$ \\ 
SM fermions& $c$&$s$&$\mu$&$\nu_{\mu}$ \\ 
& $t$&$b$&$\tau$&$\nu_{\tau}$\\
\hline
\hline new
fermions& $t^{'}$&$b^{'}$&$\tau^{'}$&$\nu_{\tau^{'}}$ \\ 
\hline 
\end{tabular}
\end{center} 
\caption{The elmentary particle spectrum of SM4}
\end{table}

 In this note, we investigate the inclusive decays of $B\rightarrow X_sl^+l^-$ and
 $B\rightarrow X_s \gamma$  in the four generation SM  which we shall call SM4 hereafter for the
sake of simplicity. These two rare $B$ meson decays 
 provide testing grounds for the SM and are very useful for constraining new physics beyond
 the SM \cite{Ali,Buras}. They are experimentally clean, and are sensitive 
 to the various extensions to the SM because these decays occur only through loops in the SM. 
 New physical effects can manifest themselves in these rare decays through the Wilson
 coefficients, which can have values distinctly different from their SM counterparts
 \cite{goto,yan,huang,wu}, as well as new operators \cite{huang}.  The implication of a 
fourth generation of quarks  on the process $b\rightarrow s$
have previously investigated \cite{wu,hh} and it is shown that the fourth generation $b\rightarrow s\gamma$
branching ratio is essentially within the range allowed by CLEO \cite{hh}.

The fourth generation quarks would influence these two decays. We
obtain two solutions of the fourth generation CKM factor $V^{*}_{t^{'}s}V_{t^{'}b}$ from
the decay of $B\rightarrow X_s \gamma$. Then we use these two solutions to calculate the new contributions of
the fourth generation quark to Wilson coefficients of the  decay of $B\rightarrow X_sl^+l^-$.  
The branching ratio and the forward-backward asymmetry of the decay of $B\rightarrow X_sl^+l^-$ in the
two cases are calculated. Our results are quite different from that of SM in one
case, almost same in another case. If Nature chooses the formmer, the $B$ meson decays could provide a
possible test for the existence of the fourth generation fermions.

\section{The decay of $B\rightarrow X_{s}\gamma$ and the fourth generation CKM factor
$V^{*}_{t^{'}s}V_{t^{'}b}$}
\label{sec:the}

 The rare decay $B\rightarrow X_s \gamma$ plays an 
 important role in present day phenomenology. The effective Hamiltonian for $B\to X_s\gamma$ at scales 
  $\mu_b={\cal O}(m_b)$ is given by \cite{Buras,cel}
    \begin{equation} \label{Heff_at_mu}
        {\cal H}_{\rm eff}(b\to s\gamma) = 
           - \frac{G_{\rm F}}{\sqrt{2}} V_{ts}^* V_{tb}
           \left[ \sum_{i=1}^6 C_i(\mu_b) Q_i + 
           C_{7\gamma}(\mu_b) Q_{7\gamma}
          +C_{8G}(\mu_b) Q_{8G} \right]\,,
    \end{equation} 
  The last two operators in the eq.(1), characteristic for this decay, are the
  magnetic--penguin operators
    \begin{equation}\label{O6B}
      Q_{7\gamma}  =  \frac{e}{8\pi^2} m_b \bar{s}_\alpha \sigma^{\mu\nu}
          (1+\gamma_5) b_\alpha F_{\mu\nu},\qquad           
       Q_{8G}     =  \frac{g}{8\pi^2} m_b \bar{s}_\alpha \sigma^{\mu\nu}
           (1+\gamma_5)T^a_{\alpha\beta} b_\beta G^a_{\mu\nu}             
    \end{equation}
  
 The leading logarithmic calculations can be summarized in a
  compact form  as follows \cite{Buras}:
    \begin{equation}\label{main}
      R_{{\rm quark}} =\frac{Br(B \to X_s \gamma)}
       {Br(B \to X_c e \bar{\nu}_e)}=
     \frac{|V_{ts}^* V_{tb}^{}|^2}{|V_{cb}|^2} 
     \frac{6 \alpha}{\pi f(z)} |C^{\rm eff}_{7}(\mu_b)|^2\,,
    \end{equation}
  where 
    \begin{equation}\label{g}
      f(z) = 1 - 8z + 8z^3 - z^4 - 12z^2 \ln z           
       \quad\mbox{with}\quad
        z =
       \frac{m^2_{c,pole}}{m^2_{b,pole}}
       \end{equation}
  is the phase space factor in $Br(B \to X_c e \bar{\nu}_e)$ and
  $\alpha=e^2/4\pi$. The coefficient $C^{\rm eff}_{7}(\mu_b)$ can be
  calculated by using the renormalization group equations and the values of the Wilson coefficients 
   $C_{7\gamma}$ and $C_{8G}$ at the scale $\mu_W=O(m_W)$,
   $C_{7\gamma}(\mu_W)$ and  $C_{8G}(\mu_W)$, 
  which in SM are given in Ref. \cite{il}.

   In the case of four generation there is an additional contribution to $B\rightarrow X_s\gamma$ 
from the virtual exchange of the 
fourth generation up quark $t^{'}$. The Wilson coefficients of the dipole operators are given by
   \begin{equation}
      C^{\rm eff}_{7,8}(\mu_b)=C^{\rm (SM)\rm eff}_{7,8}(\mu_b)
      +\frac{V^{*}_{t^{'}s}V_{t^{'}b}}{V^{*}_{ts}V_{tb}}C^{(4)
      {\rm eff}}_{7,8}(\mu_b),
    \end{equation}
 where $C^{(4){\rm eff}}_{7,8}(\mu_b)$ present the contributions of $t^{'}$ to the Wilson coefficients, and 
$V^{*}_{t^{'}s}$ and $V_{t^{'}b}$ are two elements of the $4\times 4$ CKM matrix which now contains nine
paremeters, i.e., six angles and three phases. We recall here that the CKM coefficient 
corresponding to the $t$ quark contribution, i.e., $V_{ts}^*V_{tb}$, is factorized in 
the effective Hamiltonian as given in Eq. (1). The formulas for 
calculating the Wilson coefficients $C_{7,8}^{(4)}(m_W)$ are same as their counterpaters in 
 the SM except exchanging  $t^{'}$ quark not $t$ quark and the corresponding 
 Fenymann figuers are shown in fig. 1.
 
 \begin{figure}

\epsfxsize=16cm
\epsfysize=9.0cm
\centerline{
\epsffile{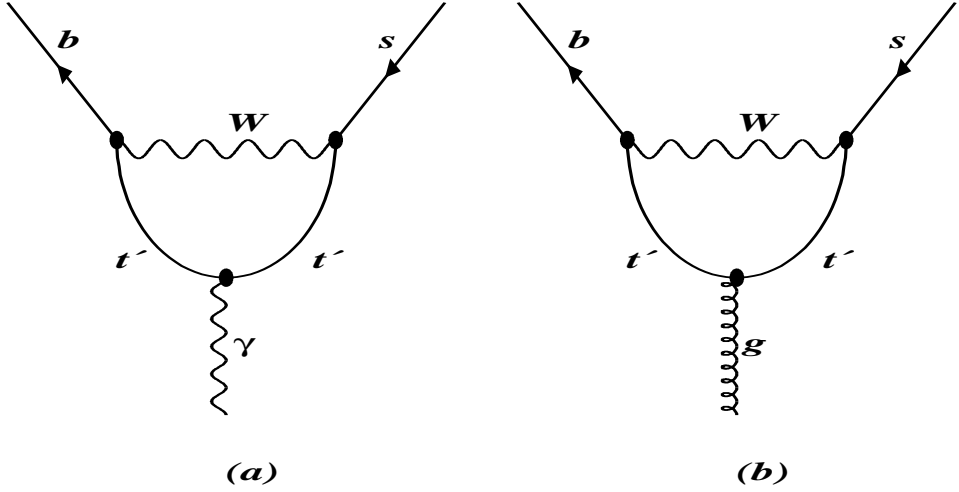}}

\caption{Mabnetic Photon (a) and Gluon (b) Penguins with $t^{'}$.}
\end{figure}

  With these Wilson coefficients and the experiment results of the decays of
 $B\rightarrow X_{s}\gamma$ and $Br(B \to X_c e \bar{\nu}_e)$ \cite{al,data}, we obtain 
 the results of the fourth generation CKM factor $V^{*}_{t^{'}s}V_{t^{'}b}$. 
 There exist two cases, a positive factor and a negative one: 
    \begin{eqnarray}
      V^{*}_{t^{'}s}V_{t^{'}b}^{(+)} &=& [C^{(0){\rm eff}}_{7}(\mu_b)
        - C^{\rm (SM)\rm eff}_{7}(\mu_b)]
       \frac{V_{ts}^* V_{tb}}{C^{(4){\rm eff}}_{7}(\mu_b)} \nonumber \\
        &=&
      [\sqrt{\frac{R_{\rm quark}|V_{cb}|^2\pi f(z)}{
        |V_{ts}^* V_{tb}|^2 6 \alpha}}-C^{\rm (SM)\rm eff}_{7}(\mu_b)]
        \frac{V_{ts}^* V_{tb}}{C^{(4){\rm eff}}_{7}(\mu_b)}
    \end{eqnarray}
    \begin{equation}
    V^{*}_{t^{'}s}V_{t^{'}b}^{(-)}=[-\sqrt{\frac{R_{\rm quark}|V_{cb}|^2\pi 
    f(z)}{|V_{ts}^* V_{tb}|^2 6 \alpha}}-C^{\rm (SM)\rm eff}_{7}(\mu_b)]
    \frac{V_{ts}^* V_{tb}}{C^{(4){\rm eff}}_{7}(\mu_b)}
   \end{equation}
as in tab. 2,

\begin{table}[htb]
\begin{center}
\begin{tabular}{|c|c|c|c|c|c|c|c|c|}
\hline
 $m_{t^{'}}$(Gev) & 50 & 100 & 150 & 200 &250 &300 &400 \\
\hline
\hline
$V^{*}_{t^{'}s}V_{t^{'}b}^{(+)}\times 10^{-2}$&$-11.591$&$-9.259$&
$-8.126$&$-7.501$&$-7.116$&$-6.861$&$-6.548$ \\
\hline
$V^{*}_{t^{'}s}V_{t^{'}b}^{(-)}\times 10^{-3}$&$3.5684$&$2.8503$&$2.5016
$&$2.3092$&$2.191$&$2.113$&$2.016$ \\
\hline
\end{tabular}
\end{center}
\caption{The values of $V^{*}_{t^{'}s}\cdot V_{t^{'}b}$ due to masses of 
$t^{'}$ for $Br(B\rightarrow X_{s}\gamma)=2.66\times 10^{-4}$}
\end{table}
In the numerical calculations we set $\mu_b=m_b=5.0GeV$  and take the 
$t^{'}$ mass value of 50GeV, 100GeV, 150GeV, 200GeV, 250GeV, 300GeV, 400 Gev
\cite{McKay}.

The CKM matrix elements obey unitarity constraints, which states that any pair of rows, or
any pair of columns, of the CKM matrix are orthogonal.  This leads to six orthogonality
conditions \cite{Ali}.  The one relevant to $b\rightarrow s\gamma$ is 
\begin{equation}
\sum\limits_{i}V_{is}^{*}V_{ib}=0, 
\end{equation} 
i.e.,
\begin{equation} V_{us}^{*}V_{ub}+V_{cs}^{*}V_{cb}+V_{ts}^{*}V_{tb}
+V_{t^{'}s}^{*}V_{t^{'}b}=0. \end{equation} 
We take the average values of the SM CKM matrix
elements  from Ref. \cite{data}.  The sum of the first three terms in eq.  (9) is
about $7.6\times 10^{-2}$.  If we take the value of $V^{*}_{t^{'}s}V_{t^{'}b}^{(+)}$ given in
Table 2, the result of the left of (9) is much better and much more close to $0$ than that in SM,  
because the value of $V^{*}_{t^{'}s}V_{t^{'}b}^{(+)}$ is very close to the sum
but has the opposite sign. If we take $V^{*}_{t^{'}s}V_{t^{'}b}^{(-)}$, the result would
change little because the values of $V^{*}_{t^{'}s}V_{t^{'}b}^{(-)}$ are about $10^{-3}$
order, ten times smaller than the sum of the first three ones in the left of (9).
Considering that the data of CKM matrix is not very accurate, we can get the error range
of the sum of these first three terms.  It is about $\pm 0.6\times 10^{-2}$, much larger
than $V^{*}_{t^{'}s}V_{t^{'}b}^{(-)}$. Thus, the values of  $V^{*}_{t^{'}s}V_{t^{'}b}$ in the
both cases satisfy the CKM matrix unitarity constraints.

\section{The decay of 
$B\rightarrow X_sl^+l^-$}
\label{sec:the decay}

The effective hamiltonian of the decay of $B\rightarrow X_sl^+l^-$ is
  \begin{eqnarray}
    H_{eff} &=& \frac{4G_{F}}{\sqrt{2}} V_{tb} V_{ts}^{*} \sum_{i=1}^{10}
      C_{i}(\mu) O_{i}(\mu)
  \end{eqnarray}
  where $O_i$  are given in Refs. \cite{Buras,Grin}. The formulas for 
  calculating the coefficients $C_i(m_{W})$ in SM can be found in 
  \cite{Buras,Grin}. Similar to Eq. (5), the Wilson coefficients $C_9$ and $C_{10}$  which, 
in addition to $C_7$, are responsible for the decay $B\rightarrow X_s l^+l^-$ can be written in SM4 as
\begin{equation}
C_{9}(\mu_b)=C^{\rm (SM)}_{9}(\mu_b)
  +\frac{V^{*}_{t^{'}s}V_{t^{'}b}}{V^{*}_{ts}V_{tb}}C^{(4)}_{9}(\mu_b)
\end{equation}
 \begin{equation} 
  C_{10}(\mu_b)=C^{\rm (SM)}_{10}(\mu_b)
  +\frac{V^{*}_{t^{'}s}V_{t^{'}b}}{V^{*}_{ts}V_{tb}}C^{(4)}_{10}(\mu_b),
\end{equation}
where $C_{9,10}^{(4)}$, the contribution of $t^{'}$ to the Wilson coefficient $C_{9,10}$, is easily
obtained by using the expression in SM with substituting $m_t$ for $m_{t^{'}} $.

The effective Hamiltonian results in the
  following matrix elements for $B \rightarrow X_{s} l^{+} l^{-}$
   \begin{eqnarray}
    M &=& \frac{G_{\bf F}\alpha}{\sqrt{2} \pi} V_{tb} V_{ts}^{*} 
      [C_9^{eff} {\bar s_L}\gamma_{\mu}b_L{\bar l}\gamma^{\mu}l + 
    C_{10} {\bar S_L}\gamma_{\mu}b_L{\bar l}\gamma^{\mu}\gamma^5l 
      \nonumber \\  
    & & +2C_7m_b{\bar s_L}i\sigma^{\mu\nu}\frac{q^{\mu}}{q^2}b_R{\bar l}
    \gamma^{\nu}l  ],
\end{eqnarray}
here these coefficients are evaluated at $\mu$=$m_b$. 
 $C_9^{eff}$ is given as\cite{Grin}:
\begin{eqnarray}
C_9^{eff} &=& C_9 +[g(\frac{m_c}{m_b},s)+\frac{3}
{\alpha^2}k\sum_{V_{i}}\frac{\pi M_{V_{i}} \Gamma(V_{i} \rightarrow l^{+} 
l^{-})}{M_{V_{i}}^2 -q^2-iM_{V_{i}}\Gamma_{V_{i}}}](3C_1 +C_2)
\end{eqnarray} 

 From eq.(13), by integrating the angle variable of the double differential 
distributions from 0 to $\pi$, the invariant dilepton mass distributions
can be calculated and given below
\begin{eqnarray}
\frac{{\rm d}\Gamma(B\rightarrow X_sl^{+}l^{-})}{{\rm d}s}
 &=& B(B\rightarrow X_c l {\bar \nu}) \frac{{\alpha}^2}
 {4 \pi^2 f(m_c/m_b)} (1-s)^2(1-\frac{4t^2}{s})^{1/2}
 \frac{|V_{tb}V_{ts}^{*}|^2}{|V_{cb}|^2} D(s), \nonumber \\
 D(s) &=& |C_9^{eff}|^2(1+\frac{2t^2}{s})(1+2s)
      + 4|C_7|^2(1+ \frac{2t^2}{s})(1+\frac{2}{s}) \nonumber \\
    & &  + |C_{10}|^2 [ ( 1 + 2s) + \frac{2t^2}{s}(1-4s)]
      +12 {\rm Re}(C_7 C_{9}^{eff*})(1+\frac{2t^2}{s})
\end{eqnarray}
where $s=q^2/m_b^2$, $t=m_{l}/m_{b}$, $B(B\rightarrow X_c l {\bar \nu})$ is
the branching ratio which takes as 0.11, $f$ is the phase-space factor expressed in eq.(4).
The forward-backward asymmetry of the lepton in the process has also been given
\begin{eqnarray}
A(s) &=&- 3 (\frac{1-4t^2}{s})^{1/2}E(s)/D(s)\nonumber\\
E(s)&=& {\rm Re}(C_9^{eff}C_{10}^{*})s 
  + 2{\rm Re}(C_7C_{10}^{*})
 \end{eqnarray}
Numerical results are shown in figs. 2 and 3.

\newcommand{\PICL}[2]
{
\begin{picture}(170,170)(0,0)
\put(0,0){
\epsfxsize=9cm
\epsfysize=9cm
\epsffile{#1} }
\put(90,0){\makebox(0,0){#2}}
\end{picture}
}

\newcommand{\PICR}[2]
{
\begin{picture}(0,0)(0,0)
\put(240,25){
\epsfxsize=9cm
\epsfysize=9cm
\epsffile{#1} }
\put(330,25){\makebox(0,0){#2}}
\end{picture}
}

\small
\newpage
\mbox{}
{\vspace{1.0cm}}

\PICL{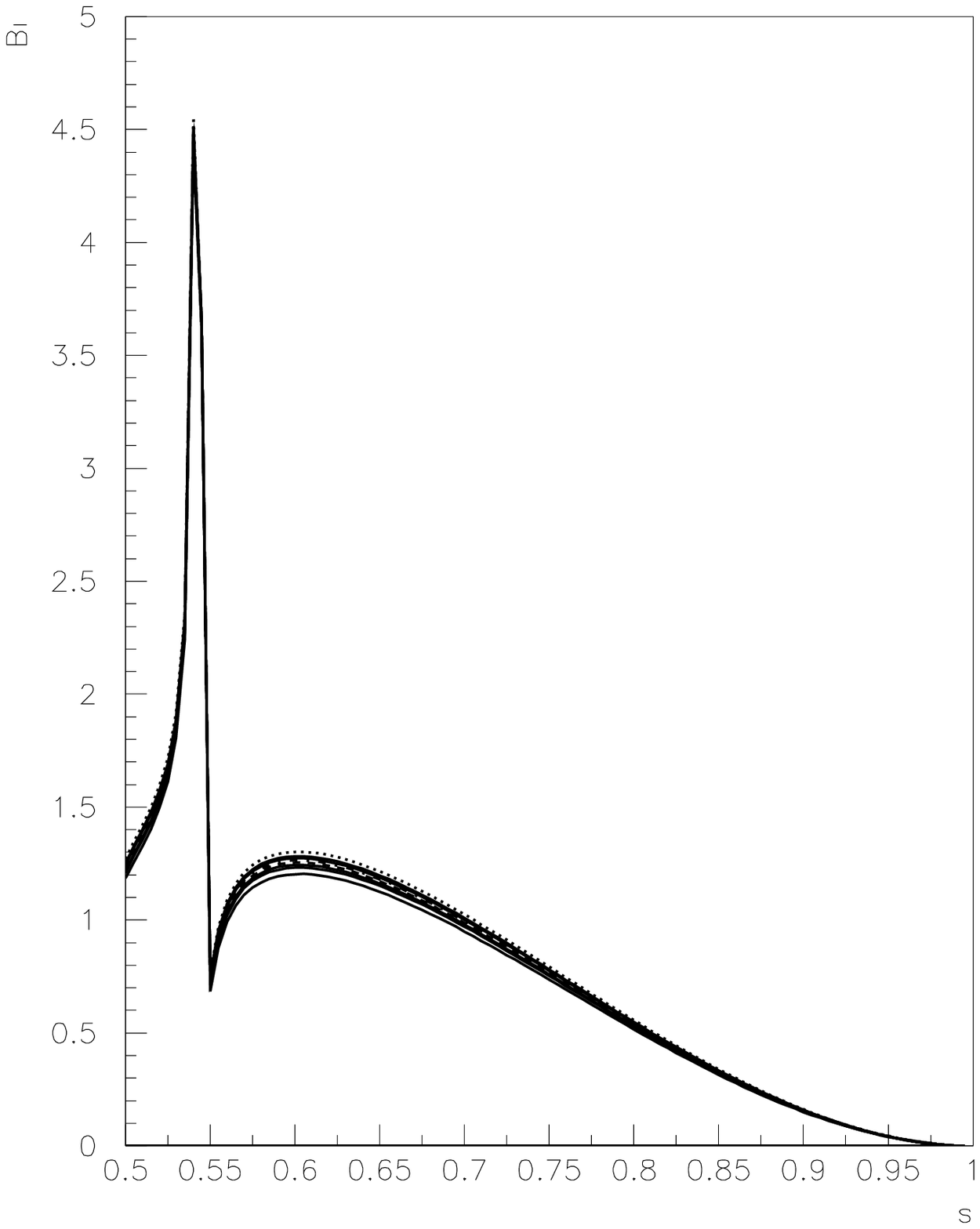}{2(a)}

\PICR{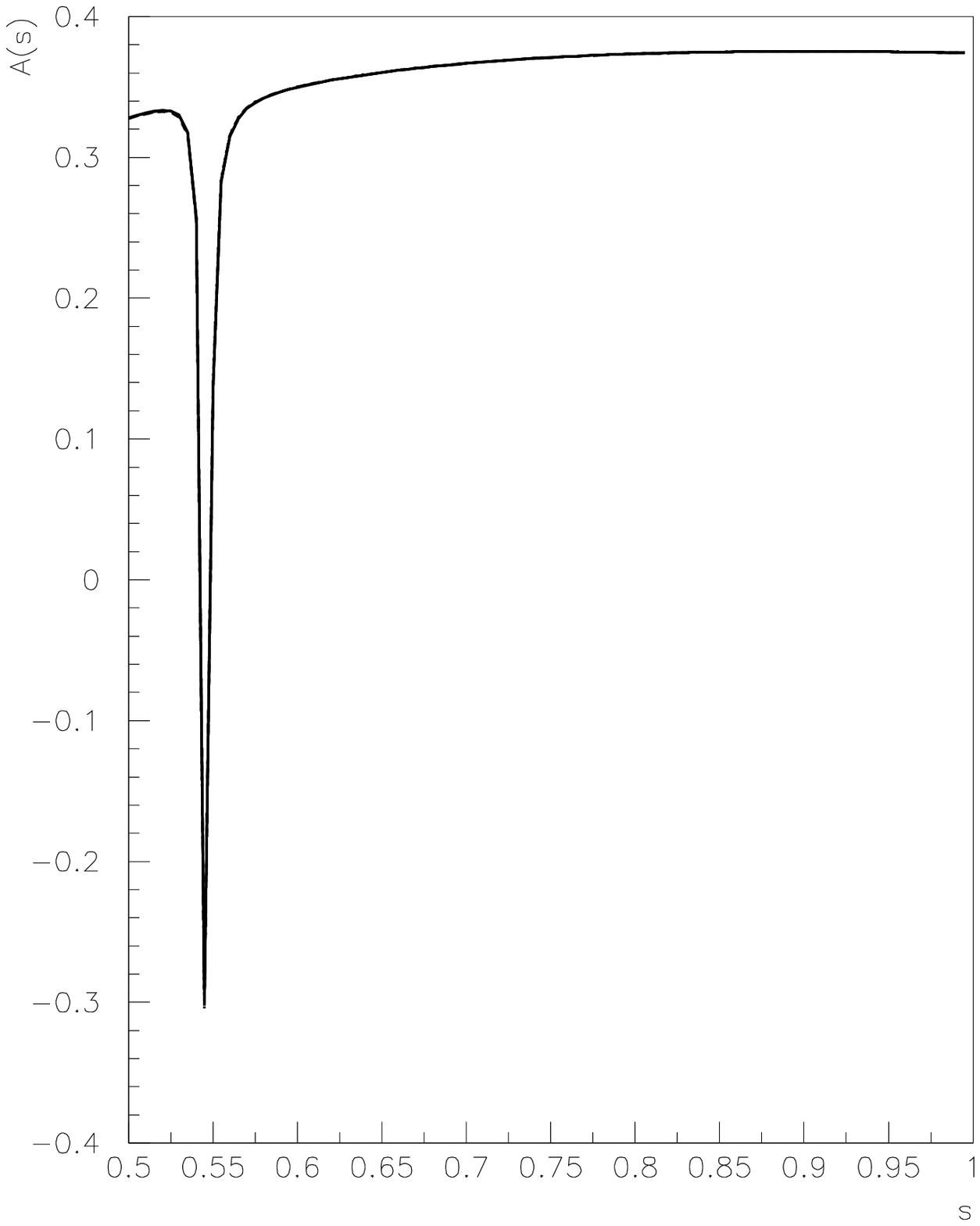}{2(b)}

{\center{Figuer 2: (a) $Br(s)$ and (b) $A(s)$ of $B\rightarrow
 X_s\tau^+\tau^-$ with massof $t^{'}$ when $V^{*}_{t^{'}s}V_{t^{'}b}$ is 
 positive.}}
\vspace{2.0cm}
\newcommand{\PICLL}[2]
{
\begin{picture}(170,170)(0,0)
\put(0,0){
\epsfxsize=9cm
\epsfysize=9cm
\epsffile{#1} }
\put(90,0){\makebox(0,0){#2}}
\end{picture}
}

\newcommand{\PICRR}[2]
{
\begin{picture}(0,0)(0,0)
\put(240,25){
\epsfxsize=9cm
\epsfysize=9cm
\epsffile{#1} }
\put(330,25){\makebox(0,0){#2}}
\end{picture}
}

\small
\mbox{}
{\vspace{1.0cm}}

\PICLL{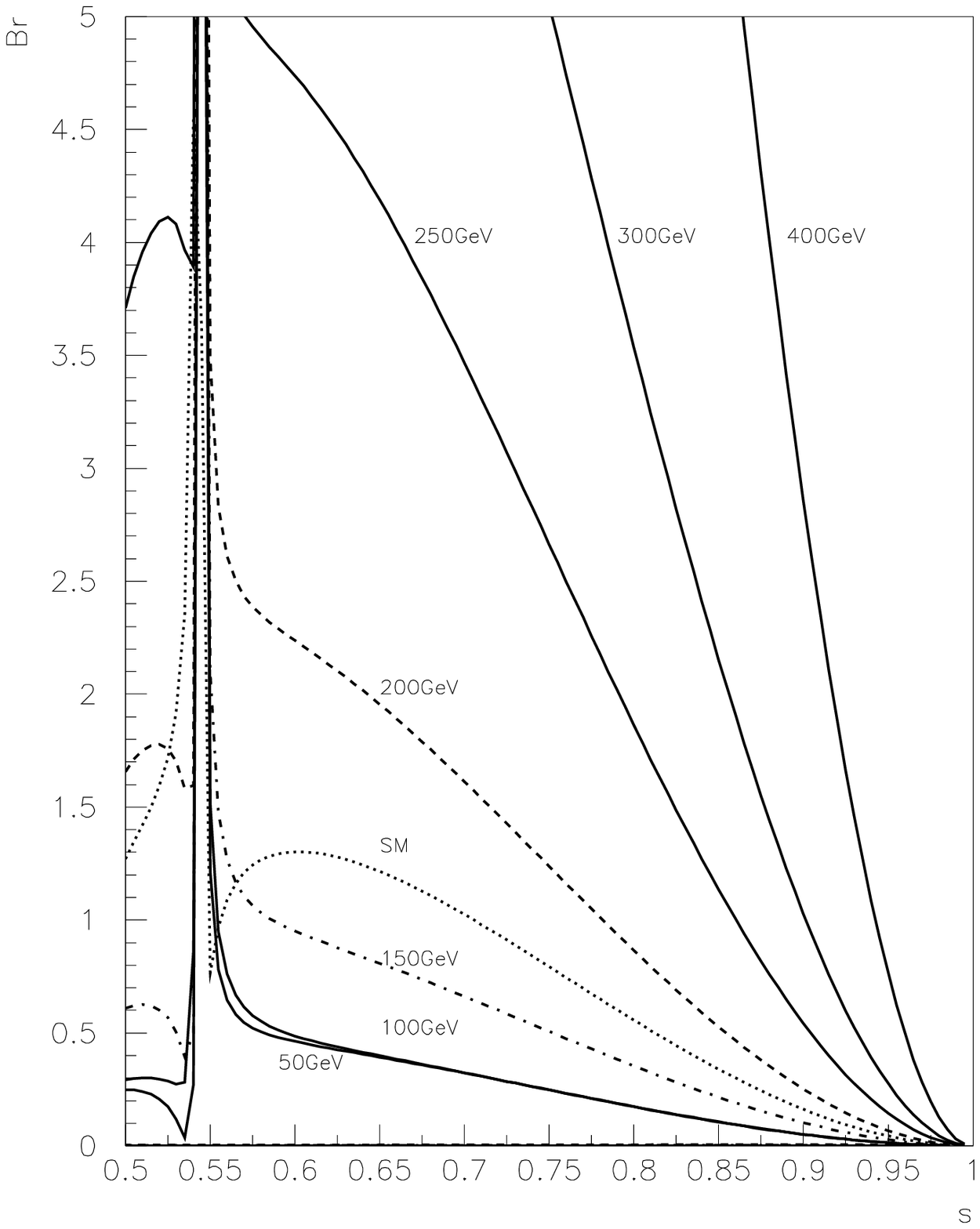}{3(a)}

\PICRR{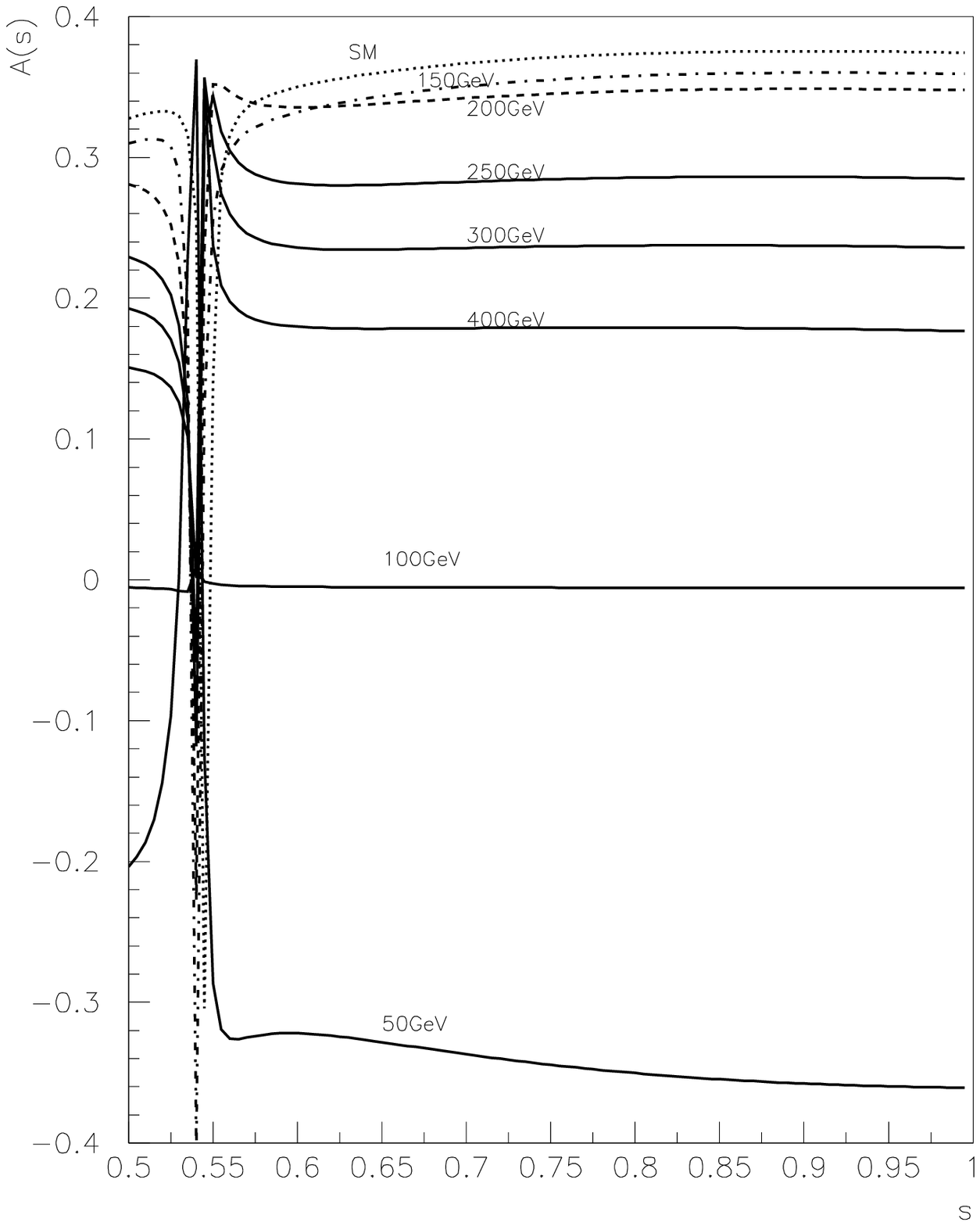}{3(b)}

{\center{ Figuer 3: same when $V^{*}_{t^{'}s}V_{t^{'}b}$ is negative.} }

The invariant mass distribution and the forward-backward asymmetry in the case of 
$V^{*}_{t^{'}s}V_{t^{'}b}^{(-)}$ are shown in the figs.2a and 2b respectively. 
The five curves, which are corresponding to $m_{t^{'}}$=50Gev, 100Gev, 150Gev, 200Gev respectively in 
four fourth generation model and the SM one, almost overlap
together.  That is, the results in SM4 are the same as that in SM.  In this case, it does not show the
new effects of $t^{'}$.  We cannot obtain the information of existence of the fourth
generation from $B$ decays, although we cannot exclude them either.  This is because, from
tab. 2, the values of $V^{*}_{t^{'}s}V_{t^{'}b}^{(-)}$ are positive. They are of order
$10^{-3}$.  The values of $V^{*}_{ts}V_{tb}$ are about ten times larger than them 
( $V^{*}_{ts}= 0.038$, $V_{tb}=0.9995$, see ref. \cite{data} ). Furthermore, $C^{(4){\rm eff}}_{7}(\mu_b)$,
$C^{(4)}_{9}(\mu_b)$, $C^{(4)}_{10}(\mu_b)$ are approximately equal to the ones in SM.  Thus the
contributions of $t^{'}$ to $C^{\rm eff}_{7}(\mu_b)$, $C_{9}(\mu_b)$,
$C_{10}(\mu_b)$ in eqs.  (5), (11), (12) are negligible .  

In the other case, when the values of $V^{*}_{t^{'}s}V_{t^{'}b}^{(+)}$ are 
 negative, the
 numerical results are very different from that of SM. This can be  clearly 
 seen from figs. 3a and 3b.  
 From fig. 3a, it is found that the deviations from SM depend on the mass of 
 $t^{'}$.  The enhancement of
 the invariant mass distribution increases with  increasing of $t^{'}$ quark
  mass.  It gets to the most
 largest deviation, about $100\%$ enhancement compared to SM, when $t^{'}$ 
 mass is 200GeV.  When the
 mass is taken to be 150GeV, the invariant mass distribution  is about 
 $30\%$ low than that in SM.  
If the mass is taken to be under 100GeV, it is about half of the SM one.  
Even we take the mass of
 $t^{'}$ as a fitting value (between 150GeV and 200GeV), the shape of the 
 curve is very
 different from the SM one. So, in this case, the fourth generation effects 
 are shown clearly.
 The backward-forward asymmetry is also different from SM and show its own 
 interesting things.  
  From fig.3b, one sees that the curves are much lower than the SM one when
 the mass of $t^{'}$ is taken to be under 100GeV. The backward-forward 
 asymmetry is almost vanishing 
as $t^{'}$ mass is equal to 100GeV.  When $t^{'}$ mass is 50GeV, it is 
completely opposite to the SM one.  
But when we take the $t^{'}$mass from 100GeV to 150GeV,
 the deviation from SM becomes smaller and smaller.  Especially, when 
 $t^{'}$ mass is near 150GeV,
 the curve is very like the SM one. But when we take $t^{'}$ mass  upper
 150GeV, the backward-forward asymmetry deviates from that of SM again.
 The deviation increases with the mass. The reason is that 
 $V^{*}_{t^{'}s}V_{t^{'}b}^{(+)}$ is 2-3 times larger than 
 $V^{*}_{ts}\dot V_{tb}$ so that the second
 term of right of the eqs.  (5), (11), (12) becomes important and it deponds on the $t^{'}$ mass strongly.  
Thus, the effect of the
 fourth generation is significant.  In this case, the decay of $B\rightarrow
 X_sl^+l^-$ could be a good probe to the existence of the fourth generation.

\section{Conclusion}
\label{sec:conclusion}

In this note, we have studied the rare B decay process $B\rightarrow X_sl^+l^-$
 as well as the decay $B\rightarrow X_s \gamma$ in SM4. We obtained two solutions of the 
fourth generation CKM factor $V^{*}_{t^{'}s}\cdot V_{t^{'}b}$ from the experimental data 
of $B\rightarrow X_s \gamma$.  We have used the two
 solutions to calculate the contributions of the fourth generation quark to Wilson coefficients of 
$B\rightarrow X_sl^+l^-$.  We have also calculated
 the branching ratio and the backward-forward asymmetry of the decay 
 $B\rightarrow X_sl^+l^-$ in the two cases.  It is found that the new results
 are quite different from that of SM when the value of the fourth generation CKM factor is
 negative, almost the same when the value is positive.  Therefore, the $B$ meson decays
 could provide a possible way to probe  the existence of the forth generation if the 
fourth generation CKM factor $V^{*}_{t^{'}s}\cdot V_{t^{'}b}$ is negative.

\section*{Acknowledgments}
This research was supported in part by the National Nature Science
Foundation of China.

\end{document}